# Error Analysis of Ia Supernova and Query on Cosmic Dark Energy

Qiuhe Peng[1](qhpeng@nju.edu.cn), Yiming Hu[1], Kun Wang[2], Yu Liang[2]

1）Nanjing University, Nanjing, China。 2） Normal University of west- China

**Abstract**： Some serious faults in error analysis of observations for SNIa have been found. Redoing the same error analysis of SNIa, by our idea, it is found that the average total observational error of SNIa is obviously greater than $0.55^m$, so we can't decide whether the universe is accelerating expansion or not.
**Keywords:** Error analysis; Ia Supernova ; Accelerating expansion of the Universe.
**PACS:** 97.60.Bw , 98.80.-k

## I. Origin of the question of "Cosmic Dark Energy"

The direct evidence of accelerating expansion of the universe is by observation of SNIa[1-3] based on the advanced Philips' relations[2,5] which are based on statistics of the absolute magnitudes at maximum luminance of SNIa related with both the width of light-curve ($\Delta m_{15}$, the decline in magnitudes 15 days after the peak luminosity) and varies of the color index (B-V) of SNIa. The distance modulus ($5\log D - 5 = \mu = m - M - A + k + \cdots$) may be calculated from the absolute magnitudes(*M*) and apparent magnitude(*m*) of SNIa. Distance modulus defined as, where D is the distance(in a unit pc). *A* is intergalactic extinction, *K* is K-correction, and "$\cdots$" includes the error caused by gravitational lensing and peculiar motion of host galaxy. Advanced Philips Relation is the most accurate method up to now. But we would like to point out that $\Delta m_{15}$ is closely related to the quantity of radioactive nuclide produced in the explosive nucleosynthesis processes, and the color index change responses to the expansion and the cooling down rate of outer aerosphere in exploding. Based on the advanced Philips' relations, SALT2[2](Spectral Adaptive Light-curve Template) is a package of the software for Type Ia Supernovae light curve fitting. Following the SALT2, UNION2[3] deal with the data of 685 SNIa. The system error of the absolute magnitudes of SNIa is found by minimizing $\chi^2$ which is the normalized quadratic sum of distance modulus residual[2,3]. As a result, they come to a conclusion that universe is accelerating expanding. And Saul Perlmutter, Brian Schmidt and Adam Reiss share the 2011 Nobel Prize in Physics for their observations that type Ia supernovae indicate that the expansion of the universe is accelerating.

However, the recent researches from 2008 to 2010 about Tycho SNR have shown The idea of single accreting white dwarf model (called as "standard model") of SNIa explosion has been negated due to recent researches[4] the Tycho SNR from 2008 to 2010. The peak luminosity of SNIa is taken as standard candles no longer due to different progenitors of SNIa . What's more, the mechanism and process of the SNIa explosion, the explosive nuclear burning and the production of radioactive nuclides in the thermonuclear explosion, the expansion and cooling down of outer aerosphere in exploding and many other physical problems have not been made sure. The researches of SNIa is still in an exploratory stage, then these absolute magnitudes based on fitting SALT2 with SNIa light-curve and spectrum[1-3] are the peak luminosities of "modeling SNIa" rather than ones of real SNIa. We will analyze the data of UNION2 compilation (it is the latest and most complete SNIa compilation) in this paper and put forward our opinion.

## II. Some serious faults in error analysis

### 1）Error in absolute magnitude

The absolute bolometric magnitudes errors at maximum light of SNIa consist of as follows: a) The intrinsic error ($\Delta M_{int}$) of the absolute magnitude at maximum luminance, in our idea, is just the half width at half-maximum (HWHM) of the statistic distribution curve of the number of SNIa with the maximum luminance, rather than the systematic error founding by using the $\chi^2$ check way[1-3]. b) Some error of $M$ originates from delivered errors caused by statistical errors of the parameters *a* and *b* in original Phillips' relation[5] or $\alpha_x, \beta$ in the advanced Philips' Relations[2,3] we call it as the delivered error, $\Delta M(a,b)$. c) Some error of $M$, $\Delta M_{Obs}$ is caused by the errors of some observational quantities of both light-curve and color index in advanced Phillips's way. The total error of absolute bolometric magnitudes at maximum light ($\Delta M_{total}$) is

$$(\Delta M_{total})^2 = (\Delta M_{int})^2 + \left(\Delta M_{max}^{(phillips)}\right)^2, \quad (\Delta M_{max}^{(phillips)})^2 = (\Delta M(a,b))^2 + (\Delta M_{Obs})^2.$$

SALT2[2] and UNION2[3] didn't give the aforementioned errors separately. In fact, they broadly !merged them into the system error caused by $\chi^2$-minimization. Advanced Philips Relation is very complex, but it is sure that the minimum of $\Delta M_{max}^{(philips)}$ is at least larger than observational apparent magnitude error, $|\Delta M_{max}^{(phillips)}| > |\delta m|$ (the observational error of apparent magnitude . Furthermore, high z SNIa is faint when observed, so its $|\delta m|$ is much larger than nearby SNIa .

### 2) The systematic error found by using $\chi^2$ check test is incorrect

However, premise of $\chi^2$ check test is the errors of distance modulus for the set of SNIa obeying a Gaussian distribution. But the set of modeling SNIa in UNION2[3] including 685 SNIa is really no Gaussian distribution . Although the average error of UNION2 which contains 685 SNIa, is $0.16^m$, over $10\%$ of total SNIa are outline of $10\sigma$. If we take a subsample including 217 SNIa with very small observational average error to do the same statistics, it is found that over $10\%$ of total SNIa are outline of $5\sigma$. Really, the critical permitted outline value for outline of $2.6\sigma$ in the standardized normal distribution is $0.805\%$ .Using the $\chi^2$ check test following SALT2[2], we find that $3.796\%$ of the data are outline of $2.6\sigma$ based on the average total observational error of the distance modulus of SNIa, $0.31^m$. Obviously, the distance modulus error deviate Gaussian distribution seriously, and it is not suitable to calculate the systematic error $\sigma_{sys}$ of SNIa by $\chi^2$ check test method. In our idea the real intrinsic error of a SNIa compilation should base on statistical distribution diagram of the number of SNIa for their absolute bolometric magnitudes (see figure 1). As we don't know the exact luminosity of high z SNIa, it is the only way to use SALT2 to get the absolute bolometric magnitudes of "modeling SNIa". The intrinsic error (or proper error) of the absolute magnitude at maximum luminosity is just the HMHW of a statistic distribution curve of the number of SNIa with the maximum luminosity. It is $\Delta M_{int} = 0.38^m$, and it is much larger than the systematic error given by $\chi^2$ check test

### III. Summary and conclusions

The average total observational errors of distance modulus is $(\Delta\mu)^2 = (\Delta M_{total})^2 + (\delta m)^2$. Using the data of SNIa and observational apparent magnitude error in UNION2, and divided intervals per $\Delta z = 0.1$, we repeat the statistics (by the same $\chi^2$ check test method) to calculate this "modeling SNIa" sample's average total observational errors of distance modulus. It is found that the average total observational error of SNIa is obviously greater than $0.55^m$ (this is much larger than $0.4^m$). So we can't decide whether the universe is accelerating expansion or not (see Fig.2). And the direct observational evidence of the idea of "dark energy" is also lost by the observational error analyses of SNIa.

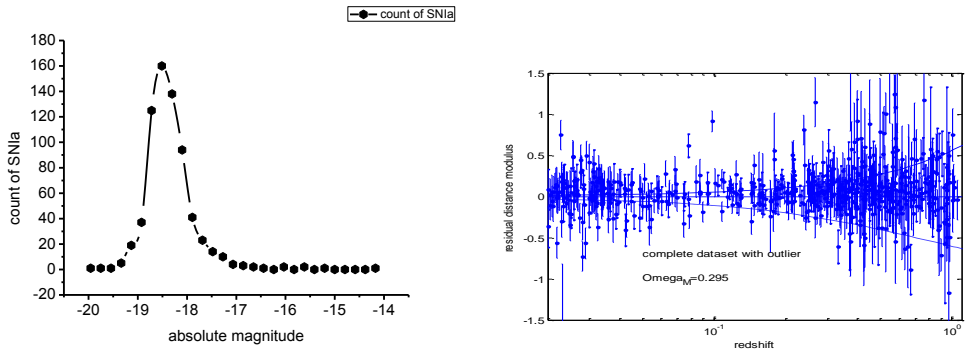